\documentclass[aps,prl,10pt,a4paper,onecolumn,showpacs,noshowkeys,notitlepage]{revtex4}

\usepackage{amsmath}
\usepackage{amssymb}
\usepackage{times}
\usepackage[dvipdfm]{graphicx}

\begin{document}

\title{Modulation of breathers in the three-dimensional nonlinear Gross-Pitaevskii equation.}

\author{A. T. Avelar$^{1}$, D. Bazeia$^{2}$, and W. B. Cardoso$^{1}$}
%\email{wesleybcardoso@gmail.com (W.B. Cardoso)}
\affiliation{$^{1}$Instituto de F\'{\i}sica, Universidade Federal de Goi\'{a}s, 74.001-970, Goi%
\^{a}nia, Goi\'{a}s, Brazil\\
$^{2}$Departamento de F\'{\i}sica, Universidade Federal da Para\'{\i}ba,
58.059-900, Jo\~{a}o Pessoa, Para\'{\i}ba, Brazil}

\begin{abstract}
In this paper we present analytical breather solutions of the three-dimensional nonlinear generalized Gross-Pitaevskii equation.
We use an \emph{Ansatz} to reduce the three-dimensional equation with space- and time-dependent coefficients into an one-dimensional
equation with constant coefficients. The key point is to show that both the space- and time-dependent coefficients of the nonlinear equation
can contribute to modulate the breather excitations. We briefly discuss the experimental feasibility of the results in Bose-Einstein condensates.
\end{abstract}

\pacs{05.45.Yv, 03.75.Lm, 42.65.Tg}
\maketitle

%%%%%%%%%%%%%%%%%%%%%%%%%%%%%%%%%%%%%%%%%%%%%%%%%%%%%%%%%%%%

\textit{Introduction} - Breathers or breathing solutions are nonlinear excitations which concentrate energy in a localized and oscillatory manner. In various physical systems, such as in Josephson junctions \cite{TriasPRL00,BinderPRL00}, charge density wave systems \cite{KleinertPRB79}, 4-methyl-pyridine crystals \cite{FillauxPRB98}, metallic nanoparticles \cite{PortalesJCP01}, conjugated polymers \cite{AdachiPRL02}, micromechanical oscillator arrays \cite{SatoPRL03}, antiferromagnetic Heisenberg chains \cite{MorisakiJPSJ07,OrignacPRB07},
and semiconductor quantum wells \cite{HassPRB09}, the breather excitations play an important role, directly affecting the electronic, magnetic, optical, vibrational and transport properties of the systems.

In the above mentioned studies, one usually considers genuine breathers, i.e., solutions which oscillate in time when the nonlinear equation presents constant coefficients (i.e., without modulation). However, in a more realistic scenario the several parameters that characterize the physical systems may depend on both space and time, leading to breather solutions that can be modulated in space and time. The presence of nonuniform and time-dependent parameters opens interesting perspectives not only from the theoretical point of view, for investigation of nonautonomous nonlinear equations, but also from the experimental point of view, for the study of the physical properties of the systems. In this context, in a recent work we have considered modulation of genuine breather solutions in cigar-shaped Bose-Einstein condensates (BECs) with potential and nonlinearity depending on both space and time, in the one-dimensional (1D) case \cite{CardosoLA09}.

The study of BECs of dilute gases of weakly interacting bosons, realized for the first time in 1995 on vapors of rubidium \cite{AndersonSCI95} and sodium \cite{DavisPRL95}, constitutes a very interesting scenario to modulate breathers, since they are well described by a three-dimensional (3D) Gross-Pitaevskii (GP) equation arising from a mean-field dynamics \cite{Pitaevskii03}. In the BEC context, one finds high experimental flexibility to control nonlinearity via Feshbach resonance, and confinement profile via optical lattices and harmonic and dipole traps \cite{Pethick02}, and there we can also investigate the effects of dimensionality reduction on the soliton solution.

In the case of a strong trapping in two spatial directions, the 3D GP equation reduces to the simpler one-dimensional (1D) form, giving rise to the so-called cigar-shaped configuration. The 1D GP equation is a nonlinear Schr\"{o}dinger equation, which can also be used to investigate pulse propagation in bulk crystals or optical fibers \cite{Kivshar03}. In a former work, however, the search for analytical solutions of the 1D GP equation with stationary inhomogeneous coefficients has been implemented via similarity transformation \cite{BronskiPRE01}. More recently, however, the case of space- and time-dependent coefficients were considered for the cubic \cite{Belmonte-BeitiaPRL08}, the cubic-quintic \cite{AvelarPRE09}, the quintic \cite{Belmonte-BeitiaPLA09}, and also the GP equation in higher dimensions \cite{GarciaPD06}.

The similarity transformation was also used to study self-similar optical pulses in competing cubic-quintic nonlinear media with distributed coefficients \cite{ZhangPRA10}, nonautonomous matter-wave solitons near the Feshbach resonance \cite{SerkinPRA10}, bright and dark solitons in a periodically attractive and repulsive potential with nonlinearities modulated in space and time \cite{CardosoNA10}, solitons of two-component Bose-Einstein condensates modulated in space and time \cite{CardosoPLA10}, and quantized quasi-two-dimensional Bose-Einstein condensates with spatially modulated nonlinearity \cite{WangPRA10}. 

On the other hand, the search for analytical solutions of the 3D GP equation has attracted a lot of attention due to the fact that solutions of higher-dimensional GP equation with constant coefficients are usually unstable \cite{Sulem99}, while the nonautonomous GP equation can engender conditions which contribute to stabilize the corresponding solutions \cite{SaitoPRL03}. Taking into account this possibility, exact solutions to 3D GP equation with varying potential and nonlinearities were proposed in Ref. 
\cite{YanPRE09}, while in Ref. \cite{YanPRA09} the authors studied analytical 3D bright solitons and soliton pairs in BECs with time-space modulation.

In this paper, our aim is to show that genuine breather solutions can be modulated by the 3D GP equation with space- and time-dependent potential, if one includes time-dependent coefficient to describe the cubic nonlinearity. In this way, we extend our recent work \cite{CardosoLA09} to the more realistic 3D case. To this end, we use an \textit{Ansatz} that changes the 3D GP equation into specific 1D equation with constant coefficients, which is easier to solve. As a consequence, however, we have to deal with a set of coupled equations, to ensure validity of the similarity transformation. Below we present explicit results for three distinct choices of potential and nonlinearity.

%%%%%%%%%%%%%%%%%%%%%%%%%%%%%%%%%%%%%%%%%%%%%%%%%%%%%%%%%%%%

\textit{Generalities} - We start with the 3D GP equation given by 
\begin{equation}
i\frac{\partial\psi}{\partial t}=-\frac{1}{2}\nabla^2\psi + v(\mathbf{r}%
,t)\psi + g(t)|\psi|^2\psi,  \label{3DNLSE}
\end{equation}
where $\psi=\psi(\mathbf{r},t)$, $\mathbf{r}\in \mathbb{R}^3$, $\mathbf{\nabla}\equiv(\partial_x,\partial_y,\partial_z)$,
and $v(\mathbf{r},t)$ and $g(t) $ are real functions representing the potential and the cubic nonlinearity, respectively.

Our goal is to find breather solutions which obey the above Eq.~(\ref{3DNLSE}). Using the 
\textit{Ansatz} 
\begin{equation}
\psi=\rho(t)e^{i\eta(\mathbf{r},t)}\Phi[\zeta(\mathbf{r},t),\tau(t)],
\label{Ansatz}
\end{equation}
we can transform the above equation into the 1D GP equation 
\begin{equation}
i\Phi_{\tau}=-\frac{1}{2}\Phi_{\zeta\zeta} + G|\Phi|^2\Phi,  \label{1DNLSE}
\end{equation}
where $\Phi_{\tau}\equiv\partial\Phi/\partial\tau$, $\Phi_{\zeta\zeta}\equiv%
\partial^2\Phi/\partial\zeta^2$, and $G$ is a constant factor. Using (\ref{Ansatz}) into (\ref{3DNLSE}) leads to (\ref{1DNLSE}), for $\rho$, $\eta$%
, $\zeta$, and $\tau$ obeying the following equations 
\begin{subequations}
\begin{eqnarray}
\tau _{t} &=& |\mathbf{\nabla}\zeta |^{2},\\%
\zeta_t + (\mathbf{\nabla}\eta)\cdot(\mathbf{\nabla}\zeta)&=&0,\\%
2\rho _{t}+\rho \nabla^2 \eta &=& 0.%
\end{eqnarray}
\label{4}
\end{subequations}
Here the potential and the nonlinearity assume the form 
\begin{equation}
v(\mathbf{r},t)=-\eta_t-|\mathbf{\nabla}\eta|^2,  \label{v}
\end{equation}
and 
\begin{equation}
g(t)=\frac{G|\mathbf{\nabla}\zeta|^2}{\rho^2}.  \label{g}
\end{equation}
Note that the potential (\ref{v}) and the nonlinearity (\ref{g}) are
functions which in general depends on the real phase $\eta(\mathbf{r},t)$ and the amplitude 
$\zeta(\mathbf{r},t)$. In this way, one can use the Eq.~(\ref{4}a) to obtain
the general form of $\zeta$, given by 
\begin{equation}
\zeta(\mathbf{r},t)=c_1(t)x+c_2(t)y+c_3(t)z+c_4(t),  \label{zeta}
\end{equation}
where the coefficients $c_j$ $(j=1,2,3,4)$ are time-dependent functions, obeying the
relationship 
\begin{equation}
\tau_t=c_1^2+c_2^2+c_3^2.  \label{tau}
\end{equation}

Now, substituting (\ref{zeta}) into (\ref{4}b) leads to $\eta$ in the general
form 
\begin{eqnarray}
\eta(\mathbf{r},t)&=& d_1 x^2 + d_2 y^2 + d_3 z^2 + d_4xy + d_5xz +d_6yz 
\notag \\
&+& d_7x + d_8y + d_9z +d_{10},  \label{eta}
\end{eqnarray}
where the $d_j$ are time-dependent coefficients which obey the equations 
\begin{subequations}
\begin{eqnarray}
&&\dot{c_1} + 2c_1d_1 + c_2d_4 + c_3d_5 =0,\\
&&\dot{c_2} + c_1d_4 + 2c_2d_2 + c_3d_6 =0,\\
&&\dot{c_3} + c_1d_5 + c_2d_6 + 2c_3d_3 =0,\\
&&\dot{c_4} + c_1d_7 + c_2d_8 + c_3d_9 =0,  \label{10d}
\end{eqnarray}
where $\dot{c_j}\equiv dc_j/dt$. Next, from (\ref{4}c) and (\ref{eta}) we obtain 
\end{subequations}
\begin{equation}
\rho(t)=\exp\left[-\int (d_1+d_2+d_3)dt \right].  \label{rho}
\end{equation}

Inserting (\ref{eta}) into (\ref{v}) leads to the potential 
\begin{eqnarray}
v(\mathbf{r},t)&=&\omega _{1}x^{2}+\omega _{2}y^{2}+\omega _{3}z^{2}+\omega
_{4}xy+\omega _{5}xz  \notag \\
&+&\omega _{6}yz +\omega _{7}x+\omega _{8}y+\omega _{9}z+\omega _{10},
\label{pot}
\end{eqnarray}%
where $\omega =\omega (t)$, with
\begin{subequations}
\begin{eqnarray}
\omega _{1}&=&\dot{d}_{1}+4d_{1}^{2}+d_{4}^{2}+d_{5}^{2}, \\
\omega _{2}&=&\dot{d}_{2}+4d_{2}^{2}+d_{4}^{2}+d_{6}^{2}, \\
\omega _{3}&=&\dot{d}_{3}+4d_{3}^{2}+d_{5}^{2}+d_{6}^{2}, \\
\omega _{4}&=&\dot{d}_{4}+4d_{1}d_{4}+4d_{2}d_{4}+2d_{5}d_{6}, \\
\omega _{5}&=&\dot{d}_{5}+4d_{1}d_{5}+4d_{3}d_{5}+2d_{4}d_{6}, \\
\omega _{6}&=&\dot{d}_{6}+4d_{2}d_{6}+4d_{3}d_{6}+2d_{4}d_{5}, \\
\omega _{7}&=&\dot{d}_{7}, \,\, \omega _{8}=\dot{d}_{8}, \,\, \omega _{9}=\dot{%
d}_{9}, \,\, \omega _{10}=\dot{d}_{10}.
\end{eqnarray}
Finally, we substitute (\ref{zeta}) and (\ref{rho}) into (\ref{g}) to get   
\end{subequations}
\begin{equation}
g(t)=G\tau _{t}\exp \left[ \int \left( d_{1}+d_{2}+d_{3}\right) dt\right].
\end{equation}

Here we note that one can get solutions through the $c_j$ coefficients, that is,
for specific choices of potential and nonlinearity, one can construct the $c_j$ functions, or, for specific choices
of $c_j$, one can get the corresponding potential and nonlinearity. We illustrate the general situation with the examples below.

%%%%%%%%%%%%%%%%%%%%%%%%%%%%%%%%%%%%%%%%%%%%%%%%%%%%%%%%%%%%

\textit{Analytical solutions} - To demonstrate the power of the method, we start
considering a breather solution of the Eq.~(\ref{1DNLSE}). A specific form of
the two-soliton breather solution is obtained for $G=-1$, which
corresponds to the explicit solution \cite{SatsumaPTPS74} 
\begin{equation}
\Phi(\zeta,\tau )=\frac{4(\cosh(3\zeta )+3e^{4i\tau }\cosh (\zeta))e^{i\tau
/2}} {(\cosh (4\zeta )+4\cosh (2\zeta )+3\cos (4\tau ))}.  \label{sol1}
\end{equation}
The binding potential of the two-soliton is equal to zero, corresponding to the unstable breather solution. 
Due to the vanishing binding potential, there is a splitting of the solution (\ref{sol1}) into two independent solitons at some point in time \cite{Malomed05}. On the other hand, the spatial and temporal modulation of the nonlinearity and the potential allows that we get stable breather solutions, since the solution given by Eq.~(\ref{Ansatz}) presents non vanishing binding potential. Indeed, as we have recently shown \cite{CardosoLA09}, the modulation of the trapping potential gives support to the coexistence of the two-soliton in the breather solution, without splitting. With this in mind, in the following we study the modulation of the breather (\ref{sol1}) in 3D spatial dimensions.

%%%%%%%%%%%%%%%%%%%%%%%%%%%%%%%%%%%%%%%%%%%%%%%%%%%%%%%%%%%%

%%%%%%%%%%%%%%%%%%%%%%%%%%%%%%%%%%%%%
\begin{figure}[bt]
\includegraphics[width=7cm]{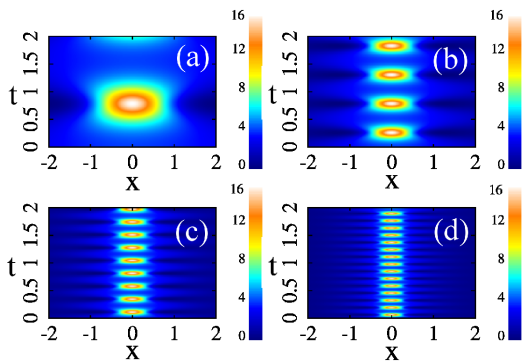}
\caption{(Color online) Plots of the 3D breather solution $|\protect\psi|^2$ for $v(\mathbf{r},t)=0$ and $g=-1$, in the $(x,t)$ plane, considering $y=z=0$. We display the
cases (a) $c_1=c_2=c_3=1/\protect\sqrt{3}$, (b) $c_1=c_2=c_3=1$, (c) $c_1=c_2=c_3=1.5$, and (d) $c_1=c_2=c_3=2$. }
\label{3df1}
\end{figure}
%%%%%%%%%%%%%%%%%%%%%%%%%%%%%%%%%%%%%

%%%%%%%%%%%%%%%%%%%%%%%%%%%%%%%%%%%%%
\begin{figure}[tb]
\includegraphics[width=4cm]{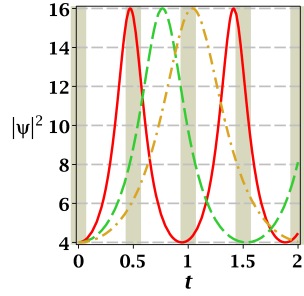}
\caption{(Color online) Plots of $|\protect\psi|^2$ for $v(\mathbf{r},t)=0$
and $g=-1$, at the spatial origin (0,0,0). We use $c_2=c_3=1/\protect\sqrt{3}$
to show the time behavior for $c_1=0.3$ in dash-dot line
(yellow), $c_1=0.6$ in dashed line (green), and $c_1=1$ in solid line (red).}
\label{3df2}
\end{figure}
%%%%%%%%%%%%%%%%%%%%%%%%%%%%%%%%%%%%%

\textit{Free evolution} - Firstly, we consider the free evolution of the
solution (\ref{sol1}) in the BEC, i.e., we take $v(\mathbf{r},t)=0$. In this way, we
can get the free evolution if we set $d_j=0$ for $j=4,5,...,10$ and $d_j=-%
\dot{c_j}/2c_j$ for $j=1,2,3$. So, we will have $\omega_j=-\ddot{c_j}/2c_j+3%
\dot{c_j}^2/2c_j^2$ with $j=1,2,3$. As an example, one gets $\omega_j=0$
setting $c_j=1/\sqrt{3}$ (and $c_4=0$), corresponding to the following choice of the
nonlinearity $g=-1$, with $\rho=1$. In this case, the Eqs.~(\ref{zeta}), (\ref{tau}),
and (\ref{eta}) change to $\zeta=(x+y+z)/\sqrt{3}$, $\tau=t$, and $\eta=0$,
respectively. Thus, we obtain the breather solution of the Eq.~(\ref{Ansatz})
in the form 
{\small 
\begin{eqnarray}  \label{i1}
&&\psi(\mathbf{r},t)=  \nonumber \\
&&\frac{4(\cosh(3(x+y+z) )+3e^{12it }\cosh (x+y+z))e^{3it /2}} {(\cosh
(4(x+y+z) )+4\cosh (2(x+y+z) )+3\cos (12t))}.  \nonumber \\
\end{eqnarray}} 
In Fig. \ref{3df1} we depict the breather solution (\ref{i1}) in the $(x,t)$ plane, considering $y=z=0$, for (a) $c_1=c_2=c_3=1/\sqrt{3}$, (b) $c_1=c_2=c_3=1$, (c) $c_1=c_2=c_3=1.5$, and (d) $c_1=c_2=c_3=2$. Similar behavior is obtained in the $(y,t)$ or $(z,t)$ plane. There we clearly see that the oscillatory frequency increases with the increasing of the values of $c_j$. We further illustrate this fact in Fig. \ref{3df2}, displaying $|\psi|^2$ at the spatial origin, considering $c_1=0.3$, $0.6$, and $1$, respectively, with $c_2=c_3=1/\sqrt{3}$.

%%%%%%%%%%%%%%%%%%%%%%%%%%%%%%%%%%%%%%%%%%%%%%%%%%%%%%%%%%%%

%%%%%%%%%%%%%%%%%%%%%%%%%%%%%%%%%%%%
\begin{figure}[tb]
\includegraphics[width=3.5cm]{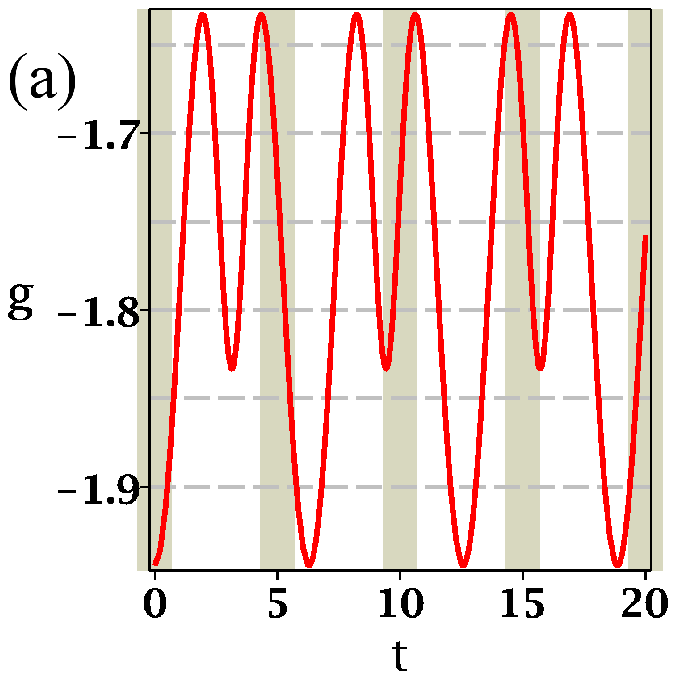}\hfil%
\includegraphics[width=3.5cm]{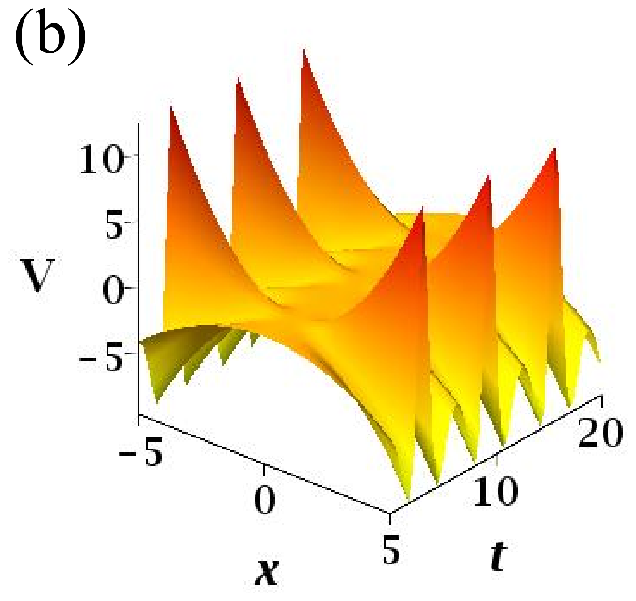}\hfil
\caption{(Color online) Plots of the nonlinearity (a) and potential (b) for
the case with $c_1=1+0.5\cos(t)$ and $c_2=c_3=1/\protect\sqrt{3}$.}
\label{np2}
\end{figure}
%%%%%%%%%%%%%%%%%%%%%%%%%%%%%%%%%%%%
%%%%%%%%%%%%%%%%%%%%%%%%%%%%%%%%%%%%
\begin{figure}[tb]
\centering
\includegraphics[width=7cm]{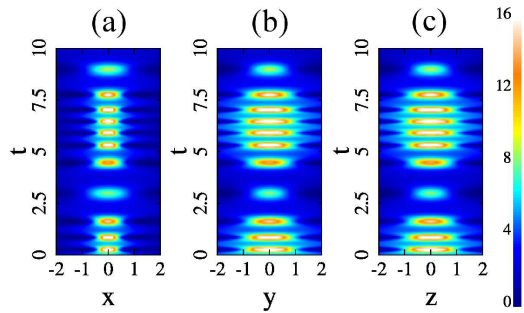}
\caption{(Color online) {Plots of the profile of the breather solution in the (a) $(x,t)$, (b) $(y,t)$, and (c) $(z,t)$ plane, in the harmonic potential for the case \textit{i}. Similar behavior appears in the case \textit{ii}.}}
\label{3df4}
\end{figure}
%%%%%%%%%%%%%%%%%%%%%%%%%%%%%%%%%%%%

\textit{Harmonic potential} - In this second example we consider the case of harmonic
potential. Firstly, we look for solutions satisfying $d_j=-\dot{c_j}/2c_j$
for $j=1,2,3$ and $d_j=0$ for $j=4,5,...10$. As before, adjusting appropriately the values of $c_j$ one can obtain the harmonic potential. We will analyze two cases: \textit{i}) harmonic
potential in a single spatial direction, say $x$, which is obtained with $c_1=1+0.5\cos(t)$ and $%
c_2=c_3=1/\sqrt{3}$; \textit{ii}) harmonic potential in the three spatial directions, which is obtained with $c_j=1+0.5\cos(t)$ for $j=1,2,3$.

\begin{figure}[tb]
\includegraphics[width=3.5cm]{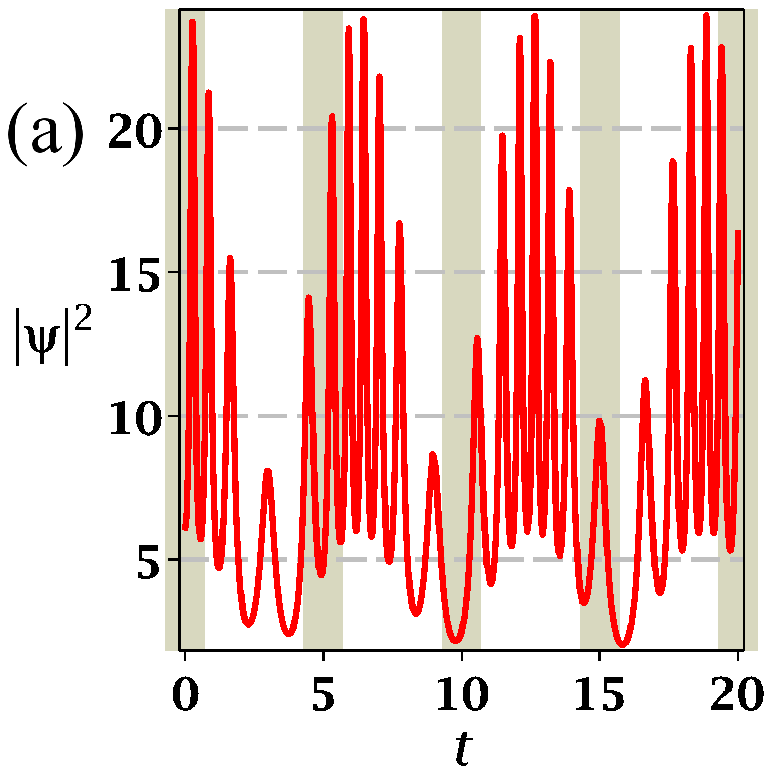}\hfil
\includegraphics[width=3.5cm]{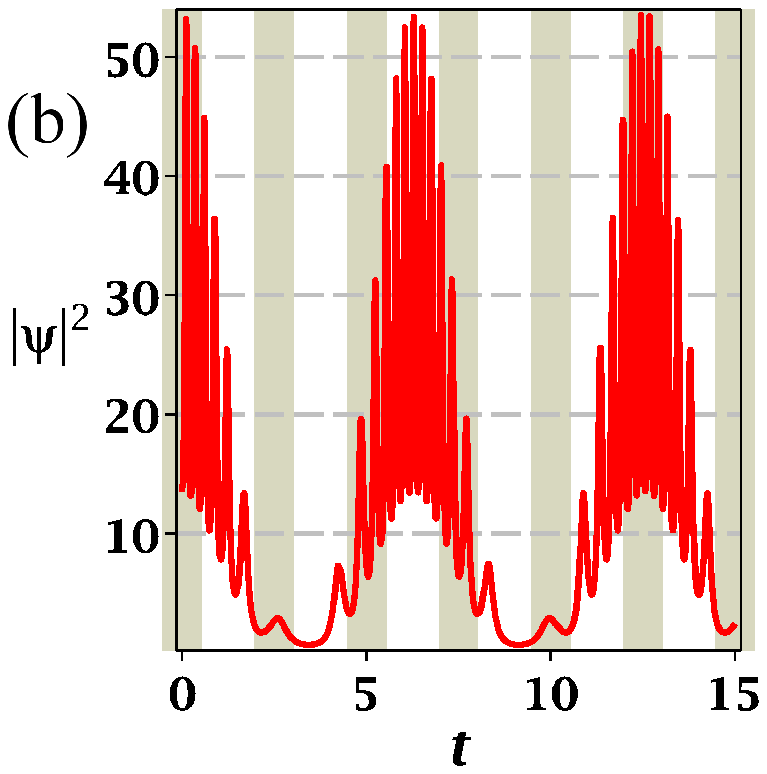}
\caption{(Color online) Profile of the breather solution for the cases (a) 
\textit{i} and (b) \textit{ii}, at the spatial origin $(0,0,0)$.}
\label{3df5}
\end{figure}

In the case \textit{i}), let us have $\tau =
1.79t+\sin(t)+\cos(t)\sin(t)/8$, $\zeta = (1+0.5\cos(t))x+\sqrt{3}y/3+\sqrt{3%
}z/3$, $\rho = \sqrt{1+0.5\cos(t)}$, $\eta = \sin(t)x^2/4(1+0.5\cos(t))$,
and $v(x,t) = -\cos(t)x^2/4(1+0.5\cos(t))-0.375\sin(t)^2x^2/(1+0.5\cos(t))^2$, which is physically implemented
by a time-dependent harmonic potential plus an optical superlattice \cite{Peil03}.
The nonlinearity and potential are displayed in Fig. \ref{np2}a and \ref{np2}b, respectively.  Also, in Fig. \ref{3df4} we depict the breather solution
for the case \textit{i}, in the plane (a) $(x,t)$, (b) $(y,t)$, and (c) $(z,t)$, respectively. Moreover, in Fig. \ref{3df5}a we plot the breather
profile at the origin. 

Next, we consider the case \textit{ii}). Here we have $\tau =
3.375t+3\sin(t)+0.375\cos(t)\sin(t)$, $\zeta =
(1+0.5\cos(t))x+(1+0.5\cos(t))y+(1+0.5\cos(t))z$, $\rho=
(1+0.5\cos(t))^(3/2)/2$, $\eta = \sin(t)[x^2+y^2+z^2]/4(1+0.5\cos(t))$, $%
g=-3/(1+0.5\cos(t))$, and $v(\bold{r},t) =
(\cos(t)^2-\cos(t)-3/2)(x^2+y^2+z^2)/(4+4\cos(t)+\cos(t)^2)$. 
This nonlinearity can be obtained experimentally in BEC by time modulated Feshbach resonance, but an optical trapping switching from red-detuned to blue-detuned laser-beam, and \textit{vice-versa}, in a periodic fashion will be required for implementation of the time-dependent potential. In Figs. \ref{3df4}a and \ref{3df5}b we depict the profile of breather solution in the $(x,t)$ plane and at the spatial origin, respectively.

%%%%%%%%%%%%%%%%%%%%%%%%%%%%%%%%%%%%
%%%%%%%%%%%%%%%%%%%%%%%%%%%%%%%%%%%%

\textit{Linear potential} - As a third example, let us consider the case in
which the BEC is trapped by a specific linear potential. Here we search for solutions of the Eq.~(\ref{pot}) with $\omega_j=0$ for $j=1,2,...,6$.
For simplicity, we also consider $\omega_{10}=0$. To this end, we choose $d_j=0$ for $j=1,2,...,6$, $d_{10}=0$, $c_1=c_2=c_3=1/\sqrt{3}$%
, $c_4=\sin(t)$, and $d_j=-\dot{c_4}/3c_j$ for $j=1,2,3$ to satisfy the Eq.~(\ref{10d}). In this case, we get $\tau=t$, $\zeta=(x+y+z)/\sqrt{3}+\sin(t)$,
$\rho=1$, $\eta=-\cos(t)(x+y+z)/\sqrt{3}$, $g=-1$, and $v(\bold{r},t)=-\sin(t)(x+y+z)/\sqrt{3}-\cos^2(t)$.

%%%%%%%%%%%%%%%%%%%%%%%%%%%%%%%%%%%%%
\begin{figure}[tb]
\centering
\includegraphics[width=5cm]{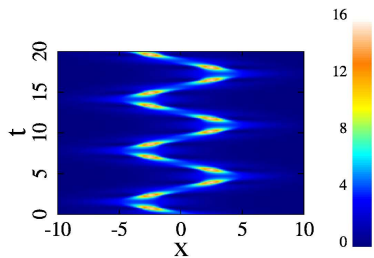}
\caption{(Color online) Plot of the breather solution in the $(x,t)$ plane, for $y=z=0$, in the presence of the linear potential. Similar behavior appears in the other $(y,t)$ and $(z,t)$ planes.}
\label{3df6}
\end{figure}
%%%%%%%%%%%%%%%%%%%%%%%%%%%%%%%%%%%%%

In Fig.~\ref{3df6} we depict the breather solution in the $(x,t)$ plane, in the presence of a linear potential. The solution presents similar behavior in the other $(y,t)$ and in the $(z,t)$ planes. Here we note that the choice $c_4=c_4(t)\neq 0$ makes the center of mass of the solution to move, as expected.

%%%%%%%%%%%%%%%%%%%%%%%%%%%%%%%%%%%%%%%%%%%%%%%%%%%%%%%%%%%%

\textit{Ending comments} - In this work we have studied the presence of breather solutions in the 3D GP equation, in the case of space- and time-dependent potential, with cubic nonlinearity described by time-dependent coefficient. We have obtained analytical solutions through an \emph{Ansatz} which changes the 3D equation into specific 1D equation.
The results show that the breather solution can be nicely modulated in space and time. We have considered three distinct examples of potential and nonlinearity: the free evolution of the breather, the case of the presence of harmonic potential and another one, in which the system is driven by linear potential. The breather solutions can be controlled through the presence of external apparatus, and this may motivate new research in the field since the modulation can generate stable excitations.

The authors would like to thank CAPES, CNPQ, and FUNAPE/GO for partial financial support.

%%%%%%%%%%%%%%%%%%%%%%%%%%%%%%%%%%%%%%%%%%%%%%%%%%%%%%%%%%%%

\end{document}